\documentclass[]{JHEP} 
\def\beq{\begin{equation}}
\def\eeq{\end{equation}}
\def\bar{\begin{eqnarray}}
\def\ear{\end{eqnarray}}
\def\id{I\kern-4.5pt I}

\usepackage{epsfig,amssymb}
\newcommand\fverb{\setbox\pippobox=\hbox\bgroup\verb}
\newcommand\fverbdo{\egroup\medskip\noindent%
            \fbox{\unhbox\pippobox}\ }
\newcommand\fverbit{\egroup\item[\fbox{\unhbox\pippobox}]}
\newbox\pippobox

\title{Solitons in 2+1 Dimensional Non-Commutative Maxwell Chern-Simons Higgs Theories }

\author{Avinash Khare\\
    Institute of Physics, Sachivalaya Marg, Bhubaneswar, 751005, India \\
    E-mail: \email{khare@iopb.res.in}}
    
\author{M. B. Paranjape\thanks{permanent address: Groupe de physique des particules, Laboratoire Ren\'e-J.-A.-L\'evesque, Universit\'e de Montr\'eal, C. P. 6128, succ. centre-ville, Montr\'eal, Qu\'ebec, Canada, H3C 3J7 }\\
    Departamento de F{\'{\i}}sica Te{\'o}rica, Facultad de Ciencias, Universidad de Zaragoza, 50009, Zaragoza, Espa\~na\\
    E-mail: \email{paranj@lps.umontreal.ca}}

\received{\today}       
\accepted{\today}       

\preprint{\hepth{0102016}}  

\abstract{
We find soliton solutions in the 2+1 dimensional non-commutative Maxwell 
Chern-Simons Higgs theories.  In the limit of the Chern-Simons coefficient going
to zero, these solutions go over to the previously found solutions in the non-commutative Maxwell Higgs theories.  The new solutions may have relevance in the theory of the fractional quantum Hall effect and 
possibly in string vacua corresponding to open strings terminating on D2 branes in the presence of D0 branes. }

\dedicated{Dedicated to\ldots\\the pursuit of knowledge.}

\begin{document} 

\maketitle 

\section{Introduction}

In the last few years, the noncommutative field theories have attracted 
considerable attention. This is primarily because such theories appear 
naturally in the various limits of M theory, as limits of 
string theories with backgrounds of open strings terminating on D2 branes, 
and are also relevant to aspects of string and D-brane dynamics 
\cite{cds}-\cite{sw}. They also appear as theories describing the behaviour 
of a two dimensional electron gas in the presence of a strong, external 
magnetic field, the quantum Hall effect \cite{suss}.  They exhibit an interesting
space uncertainty relation \cite{yon}. Being non-local, understanding of these
theories will help us in understanding the consequences of the breakdown of 
locality at short distances. 

By now, several people have looked at both the \hyphenation{per-tur-ba-tive} perturbative and 
\hyphenation{non-per-tur-ba-tive}non-perturbative aspects of the non-\hyphenation{commu-ta-tive}commutative field theories. In particular,
several people \cite{gms}-\cite{bak}, among many others, have looked at the soliton solutions in such  
theories.
For example, Jatkar et. al. \cite{jmw}, Lozano et. al. \cite{loz} as well as
Polychronakos \cite{pol} and Bak et. al. \cite{bak,blp} have looked at non-commutative $U(1)$ gauge theory
with Higgs field and have obtained the vortex (or what Polychronakos calls
as the flux tube) solutions. There is an important difference between 
the solutions obtained in \cite{pol,bak,blp} and the other two. In particular,
whereas the solutions obtained in the former are singular in the $\theta
\rightarrow 0$ limit ($\theta$ being the non-commutative parameters defined
below), the solutions obtained in the other two papers are nonsingular in this limit.

The purpose of this note is to look at the role of the Chern-Simons term on
these flux tube solutions.  In particular, we consider the non-commutative 
abelian Higgs model
with Chern-Simons term in $2+1$ dimensions and obtain soliton solutions
which represent flux tubes with both static electric and magnetic fields.  It turns out that whereas our solutions are singular in the $\theta 
\rightarrow 0$ limit, they smoothly go over to the known solutions 
\cite{pol} in the limit of the Chern-Simons coefficient going to zero.  Many authors have considered non-commutative gauge theories with a Chern-Simons term added \cite{nccs}.  The specific case of  self-dual solitons has been studied by Lozano et. al. \cite{loz1}.  

Non-commutative gauge theories are described via two equivalent descriptions.
Starting with the standard commutative gauge theory action and then 
re-interpreting any product of fields appearing in terms of the Moyal product 
is the first way. In the second approach, one re-interprets all the 
fields as operators in the Hilbert space which provides for a representation 
of the fundamental algebra that defines the underlying 
non-commutative geometry.  In this note we follow the latter approach and we will closely follow the notation of Polychronakos \cite{pol}.  

Specializing to the 2+1 dimensional case, the non-commutative geometry with 
coordinates $X$, $Y$ and $t$, and the corresponding derivative operators (which act via the adjoint action)
$\partial_X$, $\partial_Y$ and $\partial_t$ satisfy the algebra 
\bar\label{1.1}
[X,Y] =i\theta\cr
[\partial_X ,\partial_Y]={-i\over\theta}\cr
[\partial_X ,X]=[\partial_Y,Y]=1\cr
[\partial_X ,Y]=[\partial_Y,X]=0
\ear
and $t$ and $\partial_t$ commute with all of the above operators.  We take 
$\theta >0$, the case $\theta <0$ is obtained via a parity reflection.   

The algebra of the non-commutative geometry allows us to define annihilation and creation operators (note $\theta >0$)
\begin{eqnarray}
a &=& {-i\over\sqrt{2\theta}}(X +iY)\cr
a^\dagger &=& {i\over\sqrt{2\theta}}(X -iY)
\end{eqnarray}
Clearly $[a ,a^\dagger ]=1$.  The usual Fock space basis states $|k>$ are obtained via the action of the annihilation and creation operators:
$a |0>=0$, $a^\dagger |k> = \sqrt{k+1}|k+1>$ and $a|k> = \sqrt{k}|k-1>$.

\section{Action and Equations of Motion and Comments}
\subsection{Action and Equations of motion}
The action that will interest \cite{pol} us is written as 
\beq\label{2.1}
S=Tr\bigg ({1\over 4}[D_\mu ,D_\nu ][D^\mu ,D^\nu ] + {i \over 3} \lambda D^3 +
{1\over 2}[D_\mu ,\Phi ] [D^\mu , \Phi ] - e V(\Phi ) \bigg)~.
\eeq
Here $Tr =\int dt\,\, tr$, $tr$ is the trace over the associated Hilbert (Fock) space, $\lambda$ and $e$ are coupling constants, the operators $D_\mu$ are defined as
\beq
D_\mu =-i\partial_\mu + A_\mu
\eeq
and $D= dx^\mu D_\mu$ is the covariant exterior derivative operator.  The operators $D_\mu$ and $\Phi$ are hermitean operators. We define the non-commutative magnetic field by
\beq
[D_X ,D_Y] =i B.
\eeq
These covariant derivatives transform as 
\beq
D_\mu \rightarrow UD_\mu U^\dagger
\eeq
under local gauge transformations $U$.  $U$ is a unitary Hilbert space operator, hence an element of $U(\infty )$ and in general a function of $t$.
The equations of motion resulting from the action are
\beq\label{2.5}
[D_\nu ,[D^\nu ,D_\mu ] ] + [\Phi ,[D_\mu ,\Phi ] ]
=i\lambda \epsilon_{\mu\nu\lambda}[D^\nu ,D^\lambda ]~,
\eeq
\beq\label{2.6}
[D_\mu ,[ D^\mu ,\Phi ] ] + e V^\prime (\Phi ) = {0}~.
\eeq
\subsection{Comments on Non-Commutative Yang-Mills Theories, the Coefficient of the Chern-Simons Term and Gauge Invariance}
We wish to make three points.  First, we have not concerned ourselves with 
the distinction made between $U(1)$ non-commutative gauge theory and that 
based on a non-abelian group, say $U(N)$, as was done for example in 
Polychronakos \cite{pol}.  The reason is that the $U(1)$ theory contains 
all possible non-abelian theories within it, in a very precise and exact 
way (See also \cite{gn},\cite{pol},\cite{blp}).  The representation 
Hilbert space given in for example \cite{pol} for the
non-abelian case is the tensor product of a finite dimensional vector space 
taken say for the $U(N)$ factor with the standard infinite dimensional Hilbert (Fock) space for the representation of the algebra of the non-commutative geometry.  Thus a basis of this space, $\cal A$, is for example given by
\beq
{\cal A} =\{ |a)\otimes |k>,\quad a\in 1\cdots N, \quad k\in 0,1,2,\cdots \}
\eeq
and gauge transformations consist of unitary operators $V$ which act in this space.  It is evident that a relabelling of these basis states can yield a basis which appears identical to the usual Fock basis without the $N\times N$ factor.  For example if we define $\cal B$ such that
\bar
{\cal B} &=&\{ |b>>,~~ b=0,1,2,\cdots , \cr
&\ni & |b>> = |a)\otimes|k>,~~ a=b\, 
{\rm mod}\, N,~~ k={b-b\, {\rm mod}\, N\over N},~ \}
\ear
and the unitary transformations $V$ are simply unitary transformations on 
the  Fock space relabelled as in the basis $\cal B$.  Hence the 
non-commutative $SU(N)$ gauge theory is completely contained inside the 
so-called non-commutative $U(1)$ gauge theory.  Perhaps a better name for 
this gauge theory would be the $U(\infty )$ gauge theory.  We wish to make 
clear that there is still some utility to look at non-commutative Yang-Mills 
theories in the tensor product picture, indeed, we will use it later.  Our 
point is simple, that the $U(1)$ theory actually contains all possible 
gauge groups inside it. 

Secondly, the coefficient of the Chern-Simons term (as written in equation (\ref{2.1})) is imaginary for the action in Minkowski space, hence $\lambda$ is real.  This yields a real 
action for hermitean operators $D_\mu$.  Analytic continuation to Euclidean 
space results in the appearance of a relative $i$ between the Chern-Simons 
term, which is odd in temporal derivatives, and the rest of the Lagrangian 
which is even.  The ensuing Euclidean operator equations of motion have no 
solution in terms of hermitean operators $D_\mu$.  This fact is familiar in 
the usual commuting case, where Euclidean, instanton solutions for theories 
containing Chern-Simons terms exist only for complex fields, ie. complex 
monopoles \cite{hos} or complex vortex strings \cite{imbp}.  Hence the 
hermitean solution by Polychronakos \cite{pol} for the Euclidean case is not 
valid for the usual Euclidean theory, however it is a solution of this 
somewhat different theory.  

Our third, related point is that the Chern-Simons action is actually, at least formally, 
perfectly gauge invariant.  Consequently the coefficient of the Chern-Simons term, for the Euclidean theory, need not even be imaginary, relative to the other terms in the action. Furthermore, the coefficient of the Chern-Simons term need not be quantized, as one must take in the case of a commutative non-abelian gauge theory \cite{jdt}.   Hence one can actually entertain a Euclidean action and equations of motion where $\lambda$ is real, with the solution in terms of hermitean operators as given by 
Polychronakos \cite{pol}.  The analytic relationship of such a Euclidean theory and its solutions is not to the usual Minkowski theory 
with real Chern-Simons term, but one with imaginary coefficient $\lambda$.  
Obviously, such a theory has no counterpart in the usual, commutative theatre.  

\section{Solutions}

\subsection{Previously found solutions}
It is useful to rewrite the equations of motion in terms of the operators
\beq
D={D_X+iD_Y\over\sqrt 2}~,\quad 
D^\dagger ={D_X-iD_Y\over\sqrt 2} 
\eeq
which yields the equations
\beq\label{3.2}
[D,[D^\dagger ,D_0]] +[D^\dagger,[D ,D_0]]-[\Phi ,[D_0 ,\Phi ]] 
= 2\lambda [D,D^\dagger ]
\eeq
\beq\label{3.3}
[D_0 ,[D_0 ,D]] +[D,[D,D^\dagger ]] +[\Phi ,[D,\Phi ]]= 2\lambda [D_0 ,D]
\eeq
\beq\label{3.4}
[D_0 ,[D_0 ,\Phi ]]-[D,[D^\dagger ,\Phi ]] -[D^\dagger,[D ,\Phi]] 
+e V^\prime (\Phi ) = 0
\eeq
It may be noted that whereas for $\lambda =e=0$, the field eqs. 
(\ref{3.2}) to (\ref{3.4})  are scale 
invariant, the scale invariance is lost in case $e=0$ but $\lambda \ne 0$. 
In particular, for $e=\lambda =0$ the field eqs. are unchanged under 
$(D,D_0,\Phi) \rightarrow \alpha (D,D_0,\Phi)$. In fact, they are also
invariant under $(D,\Phi) \rightarrow \alpha (D,\Phi), D_0 \rightarrow
\pm \alpha D_0$.

The previously found flux tube solutions at $\lambda =0=e$ are obtained most 
efficiently via the solution generating technique of 
Harvey, Kraus and Larsen (HKL) \cite{hkl}, from the simple vacuum configuration
\beq\label{3.5}
D_0=-i\partial_t~,~~
D= \sqrt {2\over\theta }\,\, a~,~~
\Phi =\Phi_0
\eeq
with $\Phi_0=\phi_0\id$, $\phi_0$ a $c$-number and 
$\id $ is the identity operator. 
Obviously the generalization, 
to the case $e\ne 0$ with a potential of the form, for example
\beq
e V(\Phi ) =e (\Phi^2 -\Phi_0^2 )^2
\eeq
still yields a solution.  The flux tube solutions of Polychronakos and others 
are obtained, as enunciated by HKL \cite{hkl} 
via the adjoint action with the operator
\beq
S=\sum_{k=0}^\infty |k+1><k|
\eeq
which satisfies
\beq
SS^\dagger =\hat P_0 = \id - |0><0|
\eeq
while 
\beq
S^\dagger S=\id .
\eeq
Then
\begin{eqnarray}
D_\mu &\rightarrow & S^nD_\mu (S^\dagger )^n\cr
\Phi &\rightarrow & S^n \Phi (S^\dagger )^n
\end{eqnarray}
yields a solution with a flux tube of $n$ units of magnetic flux.

A minor comment is in order concerning the HKL construction.  It is  not 
necessary that the states chosen to construct the operator $S$ be the usual 
states of the Fock basis, $\{|k>, \quad k=0,1,2,\cdots \} $, created by 
the action of $a^\dagger$ on the vacuum.  Indeed, any orthonormal 
basis $\{ |j>>=\Sigma_k \alpha_{j,k}|k> \} $ can be used to construct $S$.  We will use this freedom later.
\subsection{A new solution}
We exhibit a new static solution, where the action of  $D_0$ gives $0$, of the equations for 
$\lambda =0$ and for $e=0$.  The choice above 
$\Phi =\Phi_0$ renders the commutators
\beq
[D , \Phi_0] = [D^\dagger , \Phi_0]=0.
\eeq
From the structure of the equations (\ref{3.2}) to (\ref{3.4}) 
it is clear that this is a sufficient 
condition but not a necessary one.  Another sufficient condition for $e=0$ 
is given by 
\beq
[D , \Phi ] = -([D^\dagger , \Phi ])^\dagger =\alpha\id
\eeq
where $\alpha$ is an arbitrary $c$-number, since the equations of motion now 
only include commutators of these quantities with other operators.  Hence 
we can take
\bar
D&=&\sqrt {2\over\theta } a\cr
\Phi &=&\alpha (a +a^\dagger )
\ear
which yields another, new solution that is not obtainable from the vacuum 
solution (\ref{3.5}) by 
the solution generating method of HKL.  Of course we could act with the HKL 
method on this new solution to get more solutions:
\begin{eqnarray}
D=\sqrt {2\over\theta }S^n a (S^\dagger )^n\cr
\Phi = \alpha S^n (a+a^\dagger )(S^\dagger )^n
\end{eqnarray}
For $\lambda\ne 0$ we have not found the smooth, static extension of this 
solution, but for the previously found solutions we do have such extensions, 
to which we turn next.
\subsection{Solutions with Chern-Simons term}
It suffices to find the extension of the vacuum solution  (\ref{3.5}) 
to the case 
$\lambda \ne 0$ because of the power of the method of HKL.  The solution can 
be exhibited in two ways, via a reasonably general 
hypothesis for the coefficients in a Fock space basis for the operators in 
question or derived as a relatively uncomplicated ansatz which is shown 
to satisfy the equations of motion.  We will show both forms for the solution as they are instructive.  
The solution was originally found using a non-static 
temporal gauge description.  We note that as in the commutative case, the 
Chern-Simons term induces a non-zero electric field whenever there is 
magnetic flux present.

\subsubsection{Time dependent solutions}
One way of obtaining the solution is to consider the temporal gauge, 
$A_0=0$, so that $D_0=-i\partial_t$, and to look for a time dependent 
solution.  The equations of motion become:
\beq\label{3.6}
\ddot D -[D,[D,D^\dagger ]] =2i\lambda \dot D~,
\eeq
\beq\label{3.7}
[D,\dot D^\dagger ] +[D^\dagger ,\dot D] =-2i\lambda [D,D^\dagger ]~.
\eeq
We take $\Phi =\Phi_0$ so that it decouples from the equations.
With the hypothesis
\bar
D&=&f(N) a\cr
&=&\sum_{n=0}^\infty f_{n}|n><n| a=\sum_{n=0}^\infty f_{n}\sqrt{n+1}|n><n+1|\cr
&\equiv &\sum_{n=0}^\infty g_n|n><n+1| 
\ear
we find
\beq\label{3.7a}
B=[D,D^\dagger ]=\sum_{n=0}^\infty |g_n|^2\bigg(|n><n| -|n+1><n+1|\bigg)~,
\eeq
\beq
[D,B]=\sum_{n=0}^\infty g_n(|g_{n+1}|^2-2|g_n|^2+|g_{n-1}|^2)(|n><n+1|~, 
\eeq
with the definition $g_{-1}=0$. Hence the field eqs. (\ref{3.6}) and 
(\ref{3.7}) give:
\beq\label{3.8}
\ddot g_n -g_n(|g_{n+1}|^2-2|g_n|^2+|g_{n-1}|^2)= 2i\lambda \dot g_n~,
\eeq
\beq\label{3.9}
(g_n\dot g^*_n-\dot g_ng^*_n)+2i\lambda |g_n|^2 
=(g_{n-1}\dot g^*_{n-1}-\dot g_{n-1}g^*_{n-1})+2i\lambda |g_{n-1}|^2~.
\eeq
Defining
\beq
g_n=R_{n}e^{i\gamma_n}
\eeq
the eq. (\ref{3.9}) simplifies to 
\beq\label{3.10}
2iR^2_n(-\dot\gamma_n+\lambda ) -2iR^2_{n-1}(-\dot\gamma_{n-1}+\lambda )=0~.
\eeq
which implies 
\beq
R^2_n(-\dot\gamma_n+\lambda )=0
\eeq
since $R_{-1}=0$.  If $R_n\ne 0$ we get 
\beq\label{3.10a}
\dot\gamma_n =\lambda \quad {\rm i.e.} \quad \gamma_n 
= \lambda t +\gamma_n (0)~,
\eeq
where $\gamma_n(0)$ allows for arbitrary constant phases for $g_n$, which we 
notationally suppress.  Then the imaginary part of eq. (\ref{3.8}) implies that $R_n$ is time independent and the real part yields
\beq\label{3.11}
R_{n} \bigg( R^2_{n+1} -2R^2_n+R^2_{n-1} -\lambda^2 
\bigg) =0~.
\eeq
This equation (with $\lambda =0$) has been analyzed with clarity by 
Polychronakos.  On applying his derivation to our case of $\lambda \ne 0$, 
we find that as in \cite{pol}, 
it has solutions of the form $R_{-1}=R_0=\cdots =R_{N-1}=0$ 
and the remaining $R_{n}\ne 0, n\ge N$.  We will take $N=0$ so that 
$R_0\ne 0$, the higher solutions are obtained via the method of HKL.  
The recurrence relation is easily solved
\beq\label{3.12}
R^2_n = \lambda^2 {n(n+1)\over 2} +(R^2_0)(n+1)~,
\eeq
and $R^2_0={2\over\theta}$ by continuity with the solution at $\lambda =0$.  
This gives the solution, restoring the constant phases
\beq\label{3.13}
g_n =e^{i(\lambda t+\gamma_n(0))}\sqrt{\lambda^2 {n\over 2} +{2\over\theta}}\sqrt{n+1}~,
\eeq
hence
\beq\label{3.14}
D=e^{i\lambda t}e^{i\gamma (N)}\sqrt{\lambda^2 {N\over 2} +{2\over\theta}}\,\,\, a~,
\eeq
and 
\beq\label{3.15}
B=[D,D^\dagger ]=\lambda^2 N +{2\over\theta}~.
\eeq

We can make an minor elaboration on the solutions in the sector with flux N 
for $e=0$.  Here the first $N$ $g_n$'s are zero, ie. 
$g_{-1}=g_0=\cdots =g_{n-1}=0$ and $g_n\ne 0$ for $n\ge N$.  For such a 
solution the possible configuration for $\Phi$ can be slightly more 
complicated.  Indeed the adjoint action of $S^N$ yields
\beq
\Phi =\Phi_0 (\id -\hat P_N)
\eeq
where
\beq
\hat P_N =\sum_{n=0}^{N-1} |n><n|.
\eeq
Hence $\Phi$ is proportional to the projection operator that projects on to 
the complement of the sub-space spanned by the first $N$ basis vectors.  
However, the gauge field is trivial in the sub-space spanned by the first 
$N$ basis vectors, hence $\Phi$ can actually be an arbitrary operator  
in this subspace:
\beq
\Phi = \hat P_N \Phi_N\hat P_N +\Phi_0 (\id -\hat P_N)
\eeq
where $\Phi_N$ is an absolutely arbitrary operator .

\subsubsection{Static Gauge Ansatz}

The other way of obtaining the solution is to consider static solutions.
Further, we again take $\Phi = \Phi_0$ so that it decouples from the 
field equations and we only need to solve field eqs. (\ref{3.2}) and 
(\ref{3.3}). We now now notice that the field eq. (\ref{3.2}) is trivially
satisfied if 
\beq\label{3.16}
[D_0, D] = \lambda D~.
\eeq
Further, eq. (\ref{3.3}) now requires that
\beq\label{3.17}
[D, [D,D^\dagger ]] \equiv [D, B] = \lambda^2 D~.
\eeq
We thus find that all solutions of the two eqs. (\ref{3.16}) and (\ref{3.17})
(with $\Phi = \Phi_0$) will automatically satisfy the field eqs. (\ref{3.2})
to (\ref{3.4}).  So the question boils down to finding the different solutions
to the two eqs. (\ref{3.16}) and (\ref{3.17}). Clearly, there could be 
several solutions to these two equations. 
One simplest possibility is that
\beq\label{3.17a}
[D, D^\dagger ] = - \lambda D_0~,
\eeq
so that in view of eq. (\ref{3.16}), eq. (\ref{3.17}) is automatically 
satisfied. The existence of such solutions has been mentioned by Polychronakos \cite{poly1}
where he pointed out that representations of $SU(1,1)$ provide for solutions of the equations of motion.
An illustration is obtained from the above time-dependent
solution, by transforming to the static gauge. 
If we choose
\beq\label{3.18}
D_0 = - \lambda N = -\lambda \sum_{n=0}^{\infty} n |n><n|~,
\eeq
while $D$ is as given in the time-dependent case (suppressing constant phases), i.e.
\beq\label{3.19}  
D = \sum_{n=0}^{\infty} R_n |n><n+1|~,
\eeq
then eq. (\ref{3.16}) is trivially satisfied while eq. (\ref{3.17}) is 
satisfied provided $R_n$ satisfies eq. (\ref{3.11}). Thus, we find that
the static solution is given by
\bar
D_0  & = & - \lambda N \cr
D    & = & e^{i\gamma (N)}\sqrt{\frac{\lambda^2}{2}N +\frac{2}{\theta}}~a \cr
\Phi & = \Phi_0~.
\ear
As expected, $B \equiv [D,D^\dagger ]$ being a gauge invariant quantity, 
is as previously found in the
time-dependent case and given by eq. (\ref{3.15}). One might argue that
$B \equiv [D, D^\dagger ] \ne -\lambda D_0$. However, if one redefines $D_0$ by
\beq
D_0 = -\lambda N -\frac{2}{\lambda \theta} \id ~
\eeq 
we recover \ref{3.17a}.
However, it is possible that there could be solutions for which 
eq. (\ref{3.17a}) is not valid. 

\subsubsection{More General Time-Dependent Solution}

We can generalize the time-dependent solution given above by choosing a 
higher moment ansatz for $D$ as in \cite{pol}
\beq\label{3.40}
D = f(N) a^k = \sum_{n=0}^{\infty} g(n) |n><n+k|~,
\eeq
while still assuming $\Phi = \Phi_0$. On exactly following the algebra as 
given in eqs. (\ref{3.7a}) to (\ref{3.10a}), one finds that instead of
eq. (\ref{3.11}), now $R_n$ satisfies the equation
\beq\label{3.41}
R_n \bigg ( R_{n+k}^2 -2R_n^2 +R_{n-k}^2 -\lambda^2 \bigg ) = 0~.
\eeq
with $R_{-1} =R_{-2}=...=0$. This equation decouples into $k$ equations, each
involving the coefficients $R_{kn+q}$ 
for single $q = 0,1,2,...,k-1$. On assuming
$R_{0,1,...,k-1}$ to be non-zero, the solution to the set of eqs. (\ref{3.41}) 
is easily obtained
\beq\label{3.42}
R_{nk+q}^2 = \frac{n(n+1)}{2} \lambda^2 + (n+1) R_q^2~, ~~ q = 0,1,...,k-1~.
\eeq
This solution comes about because the equations are universal and here are applied in the sense of equations with the gauge group $U(1)^k$.  It would be interesting to find solutions with the full $U(k)$ involved.
\subsection{ New Solution at $e=0$}\label{s3.4}
We can show the existence of new solutions at
both $\lambda =0$ and $\lambda \ne 0$. 
So far as we are aware, these solutions, even at $\lambda =0$ were not
known before.
We start with the 
hypothesis
\bar
\Phi &=& \sum_{m=0}^\infty r_{m} |m><m|\cr
D&=&\sum_{n=0}^\infty g_n |n><n+1|
\ear
Then
\bar
[D,\Phi ]&=&\sum_{n=0}^\infty g_n(r_{n+1}-r_{n})|n><n+1|\cr
[\Phi ,[\Phi ,D]]&=&\sum_{n=0}^\infty g_n(r_{n+1}-r_{n})^2|n><n+1|\cr
[D^\dagger ,[D,\Phi ]]&=&\sum_{n=0}^\infty |g_{n-1}|^2(r_{n}-r_{n-1})
-|g_n|^2(r_{n+1}-r_{n})|n><n|~.
\ear
As a result, the equations of motion (\ref{3.2}) to (\ref{3.4}) take the form
\beq\label{3.20}
\ddot g_n -g_n(|g_{n+1}|^2-2|g_n|^2+|g_{n-1}|^2)-g_n(r_{n+1}-r_{n})^2
= 2i\lambda \dot g_n~,
\eeq
\beq\label{3.21}
(g_n\dot g^*_n-\dot g_ng^*_n)+2i\lambda |g_n|^2
=(g_{n-1}\dot g^*_{n-1}-\dot g_{n-1}g^*_{n-1})+
2i\lambda |g_{n-1}|^2~,
\eeq
\beq\label{3.22}
-\ddot r_{n} -2(|g_{n-1}|^2(r_{n}-r_{n-1})- |g_n|^2(r_{n+1}-r_{n}))=0~.
\eeq
A solution to eqs. (\ref{3.21}) and (\ref{3.22}) 
is obtained with the hypothesis 
$\ddot r_{n}=0$ and 
$g_n=R_{n}e^{i\lambda t}$ which yields
\beq
r_{n+1} -r_{n}={c\over |g_n|^2}~,
\eeq
where $c$ is an arbitrary constant.
Consequently, eq. (\ref{3.20}) takes the form
\beq
R_{n}\bigg( R^2_{n+1} -2R^2_n+R^2_{n-1} -\lambda^2 
-{c^2\over  R_{n}^4} \bigg) =0~.
\eeq
Consider the case $R_{-1}=0$, $R_0\ne 0$.  Then we get the equation
\beq
R^2_{n+1} -R^2_n-(R^2_n-R^2_{n-1})= \lambda^2 +{c^2\over  R_{n}^4}~.
\eeq
Summing twice, first for $n=0\cdots N$ and then for $N=0\cdots M$ yields the 
equation (analogous to an integral equation)
\beq
R^2_{M+1}=\lambda^2{(M+1)(M+2)\over 2} +(M+2) (R^2_0) 
+c^2\sum_{N=0}^{M}\sum_{n=0}^{N}{1\over R^4_n}~.
\eeq
A self-consistent perturbative solution ensues for $R^2_n$ for small $c^2$.  
With $R^2_0={2\over\theta }$ we get 
$R^2_n={\lambda^2n(n+1)\over 2} +{2\over\theta }(n+1)$ for $c^2 = 0$, which 
is the same solution as found previously.  Then we can generate a 
perturbative expansion in $c^2$ by simply iterating the equation, for example 
to first order in $c^2$ we have
\beq
R^2_{M+1}=\lambda^2{(M+1)(M+2)\over 2} +{2\over\theta }(M+2) 
+c^2\sum_{N=0}^{M}\sum_{n=0}^{N}{1\over (\lambda^2{n(n+1)\over 2} 
+{2\over\theta }(n+1))^2}~.
\eeq
The latter sum is convergent and finite as $M\rightarrow\infty$ and there is no obstruction to iterating this process to generate a well defined perturbative expansion for $R_n$. 

This solution can be immediately generalized to the case of $U(1)^k$. We keep the same ansatz for
$\Phi$ but for $D$ we assume
\beq
D = \sum_{n=0}^{\infty} g(n) |n><n+k| ~,
\eeq
On following exactly the same steps as given above, a solution is obtained with
the hypothesis
$\ddot r_{n}=0$ and 
$g_n=R_{n}e^{i\lambda t}$ which now yields
\beq
r_{n+k} -r_{n}={c\over |g_n|^2}~,
\eeq
where $c$ is an arbitrary constant.
Consequently, we now have 
\beq
R_{n}\bigg( R^2_{n+k} -2R^2(n)+R^2_{n-k} -\lambda^2 
-{c^2\over  R_{n}^4} \bigg) =0~.
\eeq
We consider the case $R_{-1}=R_{2}=...=0$, and $R_{0,1,...,k-1}\ne 0$.  
Then we get the equation
\beq
R^2_{n+k} -2R^2_n+R^2_{n-k}= \lambda^2 +{c^2\over  R_{n}^4}~.
\eeq
This equation groups into $k$ uncoupled subsystems each involving the
coefficients $R_{kn+q}$ for a single $(q =0,1,2,...,k-1$). We can solve each of these $k$ equations as above.

\section{A Re-Formulation of the Non-Commutative Abelian Higgs Model}
In closing we exhibit a re-formulation of the Abelian Higgs model, where 
usually the Higgs scalar is taken to be in the fundamental representation.  
This is complementary to the usual form taken for matter fields in 
non-commutative gauge theories where, typically the adjoint action is used 
when concerning the gauge transformation properties of the matter.  It is 
easy to re-formulate matter in the fundamental representation in terms of 
real fields transforming under an adjoint action.  This is quite familiar 
in the commutative case, indeed any complex representation can be 
re-formulated as a sub-group of a larger group acting on exclusively real fields via an adjoint action.  It is somewhat more complicated in the 
non-commutative theatre, but everything does go through.  

Up to now we have found flux tube solutions in the abelian ($U(1)$) 
non-commu-ta-tive gauge theory.   These flux tubes do not resemble the usual, 
Nielsen-Olesen \cite{no} vortex type solitons familiar in the commutative 
theatre.  In the models considered so far, the Higgs field has essentially 
been a spectator, except for the one configuration where we could only 
obtain the solution implicitly in subsection (\ref{s3.4}).  To recover the 
Nielsen-Olesen type configurations we have to consider the Higgs field in 
the fundamental representation.  
However, every fundamental 
representation corresponds to the adjoint action in an appropriate, higher 
dimensional representation acting on a real (hermitean) space.  For example, 
the fundamental representation of $SU(2)$ exists inside $SO(4)$, the 
fundamental of $SU(3)$ exists inside $SO(6)$, etc.  Indeed the fundamental 
representation of $U(1)$, ie. a complex field, can be written equivalently as 
two real fields transforming via an adjoint action in $SO(2)$.  

Hence, consider the action, written exactly as before in eq. (\ref{2.1}), 
with the same equations of motion as in (\ref{2.5}) and (\ref{2.6}). 
 However, now the operators appearing in these 
equations are given by
\beq
D_\mu = -i\partial_\mu +\sigma_2 A_\mu
\eeq
while
\beq
\Phi = \Phi_1\sigma_1 +\Phi_2\sigma_3
\eeq
where $\sigma_i$ are the usual Pauli matrices.  Hence we are using the 
nomenclature of two real  fields, in  two non-commutative dimensional 
$SO(2)$ Yang Mills theory.  The $U(1)$ is explicitly extracted out of the 
$U(\infty )$ gauge symmetry of the usual non-commutative $U(1)$ gauge theory.
   Thus, although we find the notation redundant, it is still useful, as the 
equations of motion appear exactly as before.  For $\lambda = 0$ the vacuum 
solution is 
\beq
D_0=0,\quad D=\sqrt{2\over\theta }\,\, a,\quad \Phi =\Phi_0\sigma_1
\eeq
$\Phi_0$ appearing in the potential is interpreted as before  although 
now $\id =\id_{2\times 2}\otimes\id_{\rm Fock\,\, space}$.

The vortex solution of Bak \cite{bak} can be obtained by the method of HKL 
with an appropriate choice for $S$.  This choice can be found by comparing 
with Bak.  Here
\beq
\phi =\phi_1 +i\phi_2 = {1\over\sqrt{N+1}}\,\,a^\dagger
\eeq
hence
\bar
\phi_1={1\over 2}\bigg( a{1\over\sqrt{N+1}} + {1\over\sqrt{N+1}} a^\dagger\bigg) \cr
\phi_2={1\over 2i}\bigg( a{1\over\sqrt{N+1}} - {1\over\sqrt{N+1}} a^\dagger\bigg) 
\ear
This expression for the field $\phi$ in Bak \cite{bak} is exactly the 
definition of the operator $S$ of HKL
\beq
\phi = \sum_{n=0}^\infty |n+1><n|\equiv S,
\eeq
hence
\beq
\phi\phi^\dagger =SS^\dagger =\id_{\rm Fock\,\,\, space} -|0><0|
\eeq
which is a projection operator.  The construct $\Phi = \phi_1\sigma_1 
+\phi_2\sigma_3$ remarkably shares this property,
\bar
\Phi^2 &=& \phi_1^2 +\phi_2^2 +i\sigma_2[\phi_1,\phi_2]\cr
&=&\bigg( {S+S^\dagger\over 2}\bigg)^2 +\bigg( {S-S^\dagger\over 2i}\bigg)^2 +i\sigma_2\bigg[ \bigg( {S+S^\dagger\over 2}\bigg) ,\bigg( {S-S^\dagger\over 2i}\bigg) \bigg]\cr
&=&{1\over 2}(S^\dagger S +SS^\dagger )+{1\over 2}\sigma_2(S^\dagger S-SS^\dagger )\cr
&=& {1\over 2}(2\id-|0><0| -\sigma_2 |0><0|) =\id -\bigg( {1+\sigma_2\over 2}\bigg)|0><0|
\ear
which is evidently a projection operator.  We need to find the operator 
$\cal S$ such that
\beq
\Phi = \Phi_0\bigg( \bigg( {S+S^\dagger\over 2}\bigg)\sigma_1 +\bigg( {S-S^\dagger\over 2i}\bigg)\sigma_3\bigg) ={\cal S}\Phi_0\sigma_1{\cal S}^\dagger .
\eeq
It is not difficult to see that the solution for $\cal S$ is given by
\beq
{\cal S}= \sigma_++\sigma_-S
\eeq
where we choose to represent $\sigma_\pm ={1\over 2}(\sigma_1\pm i\sigma_3)$.  Then we find
\bar
{\cal SS}^\dagger &=&\sigma_+\sigma_- +\sigma_-\sigma_+SS^\dagger = \sigma_+\sigma_- +\sigma_-\sigma_+(\id_{\rm Fock\,\,\, space} - |0><0|)\cr
&=&{1-\sigma_2\over 2}+{1+\sigma_2\over 2}(\id_{\rm Fock\,\,\, space} - |0><0|)\cr
&=&\id +{1+\sigma_2\over 2}|0><0| \equiv \id -{\cal P}\cr
{\cal S}^\dagger {\cal S}&=& \id
\ear
where $\cal P$ is the appropriate projection operator, and
\bar
{\cal S}\Phi_0\sigma_1{\cal S}^\dagger &=& \Phi_0 (\sigma_-\sigma_1\sigma_-S + \sigma_+\sigma_1\sigma_+ S^\dagger )\cr
&=&\Phi_0\bigg( \bigg({1+\sigma_2\over 2}\bigg)\sigma_1S +\bigg({1-\sigma_2\over 2}\bigg)\sigma_1S^\dagger\bigg)\cr
&=&\Phi_0\bigg( \bigg({S+S^\dagger\over 2}\bigg)\sigma_1+\bigg({S-S^\dagger\over 2i}\bigg)\sigma_3\bigg)
\ear

\section{Conclusions}
In this paper we have found the extensions of the flux tube type solitons in 
non-commutative Maxwell Higgs theories to the situtation where a Chern-Simons 
term is added, among other results.  There are compelling reasons to consider 
such an extension.  Indeed, non-commutative geometry violates parity at a 
fundamental level, the commutation relations between $X$ and $Y$ break 
parity.  The resulting effective magnetic field also breaks parity.   There 
is no reason to expect that the governing action should preserve it, and hence it 
is natural to consider the inclusion of the Chern-Simons term.

In addition, the non-commutative Chern-Simons theory is proposed to describe 
the fractional quantum Hall effect \cite{suss} (see also \cite{nccs}).  The non-commutative pure Chern-Simons gauge theory is studied as a limiting low energy effective theory.  A solution of the equations of motion was shown to describe the 
fractional quantum Hall fluid quasi-particles.  The solution is conveniently 
expressed as 
\bar
y_1-iy_2&=&\sqrt {\theta\over 2}a\cr
x_1-ix_2&=&\sqrt {\theta\over 2}b
\ear
where $b$ is the operator defined by
\bar
b^\dagger |n>&=&\sqrt{n+1+\nu }|n+1>\cr
b|n>&=&\sqrt{n+\nu }|n-1>\cr
b|0>&=&0
\ear
where $x_i$ are the coordinates of the fluid particles (Eulerian coordinates) while the $y_i$ are the co-moving coordinates of the fluid particles (Lagrangian coordinates) (see \cite{suss} for details) in the continuum limit and $\nu$ is the filling factor.  

Our solution in the limit of large 
parameter $\lambda$ becomes
\beq\label{ll}
D_0=-\lambda N,\quad D=\lambda\sqrt{N\over 2}a, \quad \Phi=\Phi_0
\eeq
which does not seem to reduce to the solution of Susskind \cite{suss}.  It 
would be interesting to see the import of our solutions, especially with 
flux-tubes for the quantum Hall system.  For example, the solution obtained 
from the above solution by conjugating with $S$ and $S^\dagger$ is
\beq
D_0=-\lambda (N-\id +|0><0|),\quad D={\lambda\over \sqrt 2}{\sqrt{N(N-1)\over N+1}}\,\, a,\quad \Phi =\Phi_0 (\id -|0><0|).
\eeq
The energy of this configuration is given by
\beq
E=tr \bigg({1\over 2}(B^2 +[D^\dagger ,D_0][D_0 , D]) +e V(\Phi)\bigg).
\eeq
Normalizing so that the energy of the vacuum configuration, equation (\ref{ll}), is zero, we find that only the inhomogeneous potential 
contributes, taking for example $eV(\Phi )=e(\Phi^2 -\Phi_0^2)^2$ gives
\beq
E=tr (e\Phi_0^4 |0><0|)=e\phi_0^4.
\eeq

\acknowledgments
We thank the the organizers of Strings 2001, Mumbai, where this paper germinated, for a superbly run, first class conference and for their seductive proselytization of ordinary field theorists to string theory.  M.P. thanks the Institute of Physics, Bhubaneswar, for kind hospitality, where essentially all of this work was done. We thank A. P. Polychronakos for useful correspondence.  We also thank NSERC of Canada for financial support.

\end{document}